\documentclass[reprint,aps,pra]{revtex4-2}

\usepackage{graphicx}
\usepackage{dcolumn}
\usepackage{bm}
\usepackage{booktabs} 

\usepackage{subfig}

\usepackage{orcidlink} 
\usepackage{multirow} 
\usepackage{xcolor}
\draft 

\newcommand{\ket}[1]{$|\,#1\rangle$}
\usepackage{ulem}

\begin{document}

\preprint{APS/123-QED}

\title{Deployed quantum key distribution network: further, longer and more users}
\author{
Nathan Lecaron\textsuperscript{1}\orcidlink{0009-0001-3805-1618},
Yoann Pelet\textsuperscript{1}\orcidlink{0009-0007-9133-0337},
Grégory Sauder\textsuperscript{1}\orcidlink{0000-0001-8423-547X}, Nils Raymond\textsuperscript{2}\orcidlink{0000-0001-7321-2504}, Julien Chabé\textsuperscript{2}\orcidlink{0000-0001-5514-0690}, Clément Courde\textsuperscript{2}\orcidlink{0000-0001-7321-2504},
Anthony Martin\textsuperscript{1}\orcidlink{0000-0003-1664-2721},
Sébastien Tanzilli\textsuperscript{1}\orcidlink{0000-0003-4030-5821},
Olivier Alibart\textsuperscript{1}\orcidlink{0000-0003-4404-4067}
}

\email{olivier.alibart@univ-cotedazur.fr}

\affiliation{\textsuperscript{1}Université Côte d'Azur, CNRS, Institut de Physique de Nice, 17 Rue Julien Lauprêtre, 06200 Nice, France}
\affiliation{\textsuperscript{2}Université Côte d'Azur, CNRS, GéoAzur, 2130 Route de l’Observatoire, 06460 Caussols, France}

\date{\today}

\begin{abstract}{
Entanglement-based quantum links are intended to become the backbone of future quantum networks, enabling secure communication between distant cities. Implementing such networks requires addressing multiple practical challenges: time synchronization, interferometric system stabilization, dispersion, and loss management, as well as reliable automation. In this quest for practical QKD systems, we report the continuous exploitation of an operational link for more than 325\,hours over 50\,km between two remote locations, demonstrating the feasibility of long-duration key generation. We further extend secure key distribution up to a 100\,km operational link connecting several campuses of the University Côte d'Azur to an optical ground station, a setup compatible with future quantum satellite connections. Finally, by employing a dense wavelength-division demultiplexing strategy to separate paired photons, we enable QKD across multiple standard channels, achieving secure key exchange via the BBM92 protocol and time–energy observables in a multiple-user scenario.}
\end{abstract}

\maketitle

\section{Introduction}
In addition to proof-of-concept demonstrations, quantum technologies are currently experiencing a transition towards concrete applications. Quantum communication is driven by advances in quantum optics, integrated photonics, and high-efficiency single-photon detection. Entanglement, once regarded as a mere laboratory curiosity, has progressively emerged as a viable resource for the establishment of quantum networks (QNs). These networks, based on the controlled distribution of entanglement at a distance, are set to play a central role in many fields, from secure quantum communication~\cite{QKDreview} to the interconnection of quantum processors~\cite{Qcomputer} and sensors~\cite{Qsensing}, as well as fundamental tests of physics~\cite{fundamental}.

Quantum key distribution (QKD) stands as a mature application that has developed as a concrete use case for entanglement~\cite{bennett1992quantum}. In this respect, in order to extend the application framework and envision networks genuinely interconnected, with the capacity to distribute entanglement over long distances, numerous challenges must be overcome. In particular, increasing the number of users, improving their time synchronization and refining the phase stabilization of interferometer-based analyzers are key challenges. In addition, one shouldn't overlook improving the automation of the device, i.e. its capability to recover from failures in order to achieve continuous generation of secret keys over long periods without human intervention. Such objectives represent significant obstacles to be cleared in order to transition from controlled experimental contexts to realistic and scalable implementations.

Many previous works have addressed specific aspects individually, but none has provided an exhaustive demonstration. For instance, distance has been optimized in laboratory to reach 400\,km in Ref.~\cite{QKD_404km} while the number of connected users has reached 8 in Ref.~\cite{koduru_trusted_2020} for local demonstration, simplifying the remote synchronization and stabilization of analyzers. When considering remote demonstration, synchronization and phase-stabilization of the analyzers are mostly delegated to the classical layer of the network, and require temporary interruptions of the quantum communication as in Ref.~\cite{du2024longdistancequantumcapableinternettestbed,travagnin2019quantum,neumann2022continuous,Qotham_24} until correct resynchronization or analyzers stabilization procedure is performed. On the other hand, we believe that efficient quantum networking is achievable while these tasks are transferred from classical to quantum layer by exploiting quantum signals themselves, therefore simplifying the deployment of quantum channels coexisting with classical infrastructures.

To illustrate this statement, we present the complete automation of an energy-time entanglement-based quantum link, allowing a continuous and optimal generation of secret keys over long periods of time without any additional signal dedicated to synchronization and/or stabilization of analyzers. We build up upon our previous 50\,km QKD demonstration~\cite{pelet2023operational} and we experimentally address, in this work, distances of up to 100\,km, operating time of 300\,h and up to 36 users. The ability to perform entanglement-based QKD over a distance of 100\,km represents a critical milestone because it marks the transition from local metropolitan networks to larger-scale inter-city links. At such a distance, significant optical losses are observed, which serve to assess the robustness of our solutions and act as a catalyst for the advancement of long-distance architectures through satellites, which are required for intercontinental implementations~\cite{deForgesdeParny2023}.

More specifically, we implement a QKD protocol based on BBM92 and demonstrate secret key generation for more than 325\,hours during which no human intervention was required. We also demonstrate its operability over 18 pairs of International Telecommunication Union (ITU) channels and for up to 56\,dB of total transmission losses with two practical real-field configurations over 50\,km and 100\,km of commercial grade fiber.

This result is made possible thanks to the implementation of a high-brightness entanglement source coupled to low-loss, high-stability quantum state analyzers and high-efficiency single photon detectors. The demonstration is also enabled thanks to a custom-built post-processing software that is continuously monitoring various parameters, such as quantum bit error rates (QBERs), transmission losses, and time drift.
These metrics can each be linked to underlying physical parameters of the system (i.e. interferometer relative phase, polarization reference frame, and time synchronization), which can be continuously optimized within the QKD protocol to extract the highest achievable key rate. It also performs in real time all the post-processing operations, from sifting to privacy amplification, required to convert raw keys into secret keys.
Lastly, we interface the program with our single-photon detectors, allowing it to automatically shut down and restart during downtimes or after a critical failure of the QKD, therefore removing the need for an operator during the key sharing process.

\section{Experimental setup}

\begin{figure}[h]
    \centering
    \includegraphics[width=0.95\linewidth]{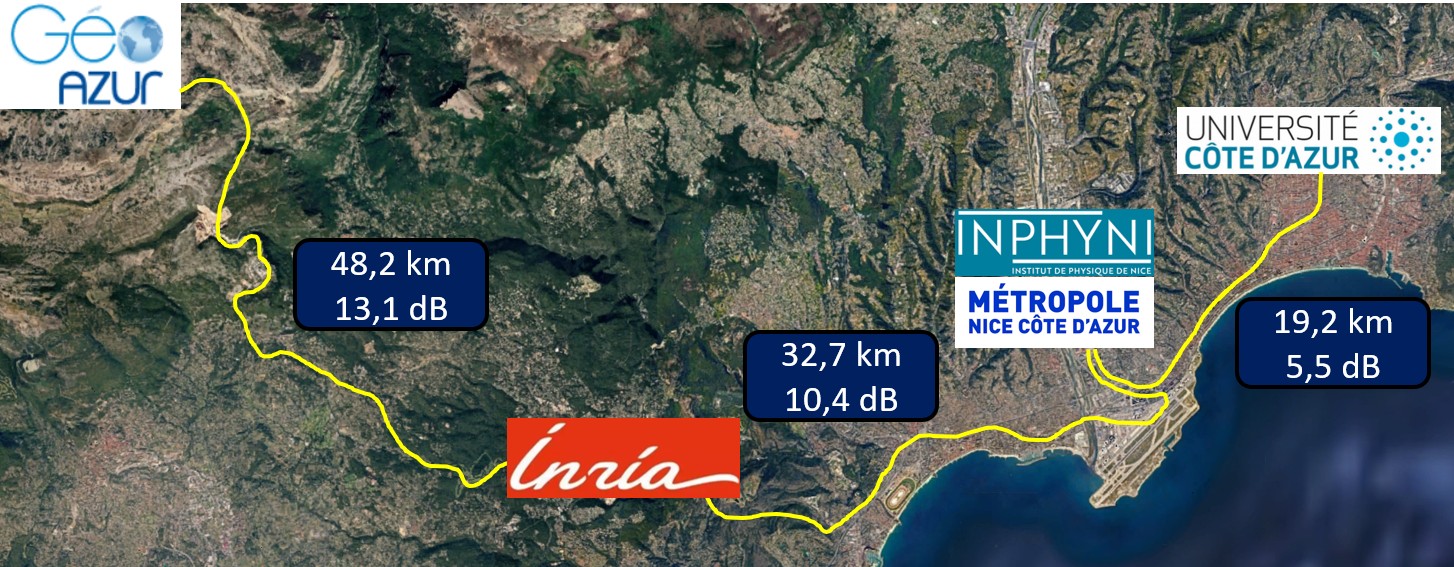}
    \caption{\centering Satellite image of the 4-nodes Côte d’Azur quantum network. For the 100\,km demonstration, the entangled photon pair source (EPPS) is located at Inria in Sophia Antipolis while Alice and Bob nodes are located at the University Côte d'Azur campus in downtown Nice, 51.9\,km from the source, and at GéoAzur in Caussols (Observatoire de la Côte d’Azur), 48.2\,km from the source, respectively. For the 50\,km demonstration, the entangled photon pair source (EPPS) is located at INPHYNI in Nice while Alice and Bob nodes are located at the University Côte d'Azur campus in downtown Nice, 19.2\,km from the source, and at Inria in Sophia-Antipolis, 32.7\,km from the source, respectively.}
    \label{qauca}
\end{figure}

More details about the fiber testbed and QKD experimental setup can be found in Ref.~\cite{pelet2023operational}, but we remind here the main features.
The newly extended setup, shown in FIG.~\ref{qauca}, is implemented over four nodes over $\sim$100\,km of deployed commercial SMF-28 optical fiber spanning across the Métropole Côte d'Azur. The network is composed of three sections of 19.2\,km, 32.7\,km, and 48.2\,km of fiber.

\begin{figure}[h]
    \centering
    \includegraphics[width=1\linewidth]{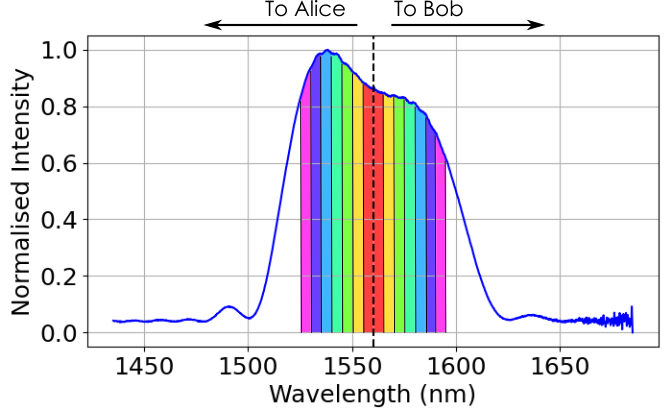}
    \caption{\centering Measured Spontaneous Parametric Down-Conversion (SPDC) emission spectrum at the output of the nonlinear crystal, centered at 1560.20\,nm. The solid blue curve shows the experimental spectrum, while the colored bands (not to scale) indicate the DWDM channels routing to Alice and Bob. Each user can measure quantum correlations inside channels of similar color.}
    \label{spdc_spectre}
\end{figure}

The energy-time entangled photon pairs generated by the source are deterministically separated according to their wavelength via spectral demultiplexing. As conceptualized in FIG.~\ref{spdc_spectre}, short wavelengths are sent to Alice and long wavelengths to Bob. Routing of the selected wavelengths can be performed actively, as in Ref.~\cite{WSS}, but we choose here to rely on low-loss passive components and exploit 18\,nm-wide Coarse Wavelength Division Multiplexing (CWDM) scheme, centered at 1531\,nm, 1551\,nm, 1571\,nm, and 1591\,nm.
With the pair emission spectrum centered at $1560.20$\,nm, we perform wavelength routing so that each user receives either the short or long wavelength of the correlated photons, as presented in more details in FIG.~\ref{fig_wavelength}. 
At both end stations, the incoming optical signals pass through Dense Wavelength Division Multiplexing (DWDM) modules, which demultiplex the incoming spectrum into up to 36 100\,GHz-channels along the standard ITU grid. 

Once separated, photons from a correlated DWDM pair are locally measured in one of two orthogonal bases (Z and X), corresponding to our observables (time and energy).
In each of those, we can compute the QBER, which provides an indirect indication of the amount of information that an eavesdropper might have access to regarding the key.
We define \textbf{the Z basis} as the time basis, in which each incoming photon can take one of two paths: short (\ket{s}) and long (\ket{l}), separated by a time delay $\tau$.
This measurement allows the generation of the sifted key by encoding a $0$ for each "short" detection, and $1$ for the "long".
\textbf{The X basis} on the other hand does not generate bits for the secret key, but instead allows us to monitor the security of the link.

Each photon is sent through an actively unbalanced Mach-Zehnder interferometer (UMZI) with a Free Spectral Range (FSR) of $2.5$\,GHz in a deployed Franson configuration~\cite{franson1989bell}, allowing us to measure locally the superposed states \ket{+}=$\frac{1}{\sqrt{2}}$(\ket{s}+\ket{l}) and \ket{-}=$\frac{1}{\sqrt{2}}$(\ket{s}-\ket{l}).
However, to get an optimal measurement, the difference between the delay of each interferometer has to be much lower than the coherence time of the photons, given by the spectral bandwidth of our filters.
In our implementation, the interferometers have a difference of $50$\,nm, while the coherence length of the photons is $\sim$3\,mm with the $100$\,GHz DWDM, allowing us to extract a minimum visibility of $99.7$\,\%.

\begin{figure}[h]
    \centering
    \includegraphics[width=1\linewidth]{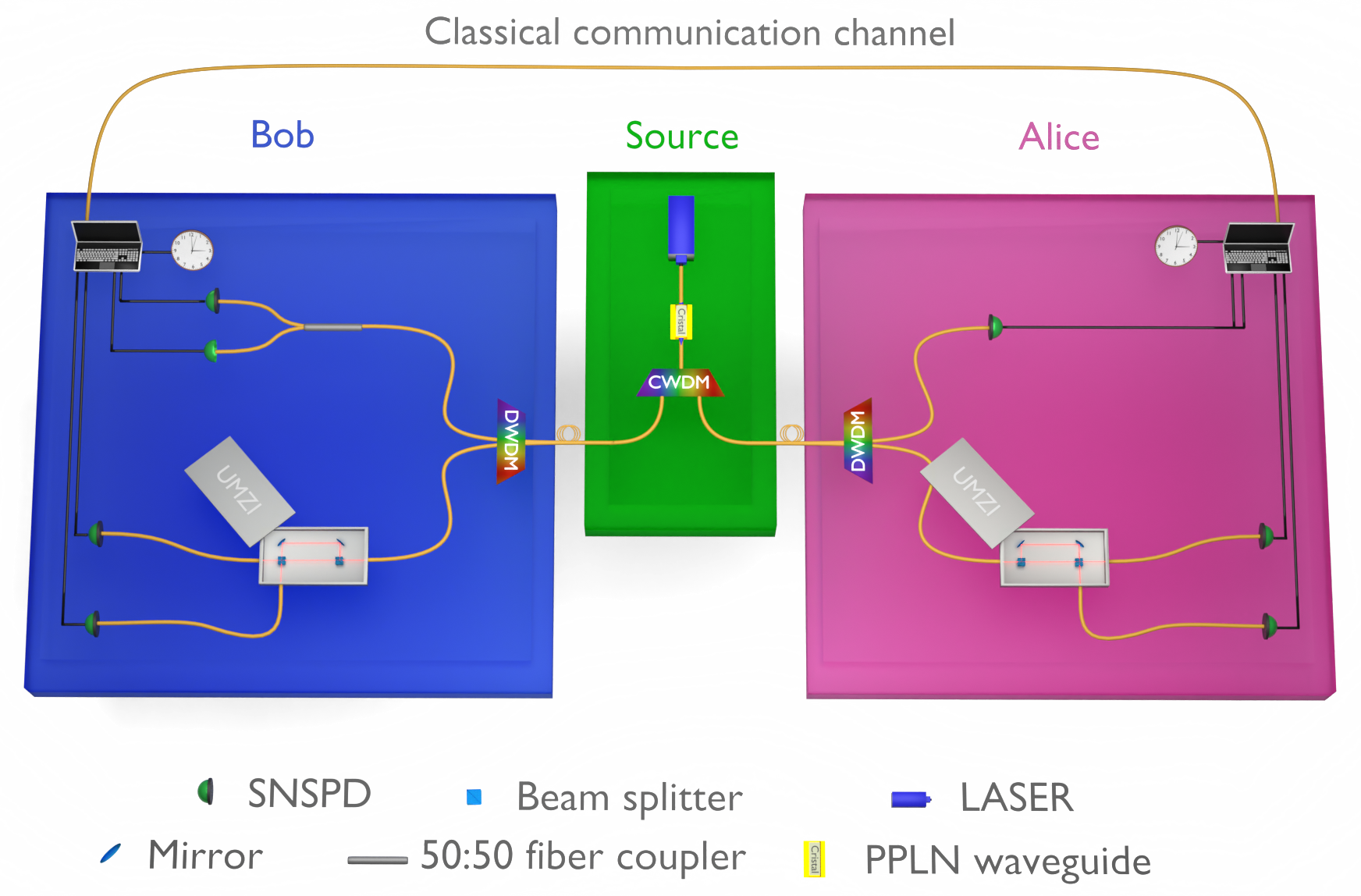}
    \caption{\centering Schematic representation of the QKD link. The source (in green) is composed of a 780\,nm CW laser, pumping a PPLN crystal, generating energy-time entangled photon pairs via Spontaneous Parametric Down-Conversion. On Alice's (pink) and Bob's (blue) stations, a Dense Wavelength Division Multiplexing (DWDM) allows to select correlated $100$\,GHz channels $20$ \& $22$, then a 50/50 beam splitter is used to passively select between Z or X basis. The former is composed of two paths, leading to two different detection times to project onto the short (\ket{s}) and long (\ket{l}) states, while the latter uses actively stabilized Michelson interferometers to project on \ket{-}=$\frac{1}{\sqrt{2}}$(\ket{s}-\ket{l}) and \ket{+}=$\frac{1}{\sqrt{2}}$(\ket{s}+\ket{l}) states.}
    \label{setup}
\end{figure}

All measurements are performed with superconducting nanowire single-photon detectors (SNSPDs) ID281 from IdQuantique, and the detection times are processed with time-to-digital converters (TDCs - Swabian Instrument Timetagger Ultra) placed at each end station. Those time-taggers are connected to rubidium atomic clocks (Spectratime LNRClock 1500) to improve the stability of their local time references.
Raw data extracted from the TDCs are sent to our in-house post-processing software for real-time secret-key generation.

\section{Continuous operation of the network}

\begin{figure}[h]
    \begin{center}
        \includegraphics[width=0.85\linewidth]{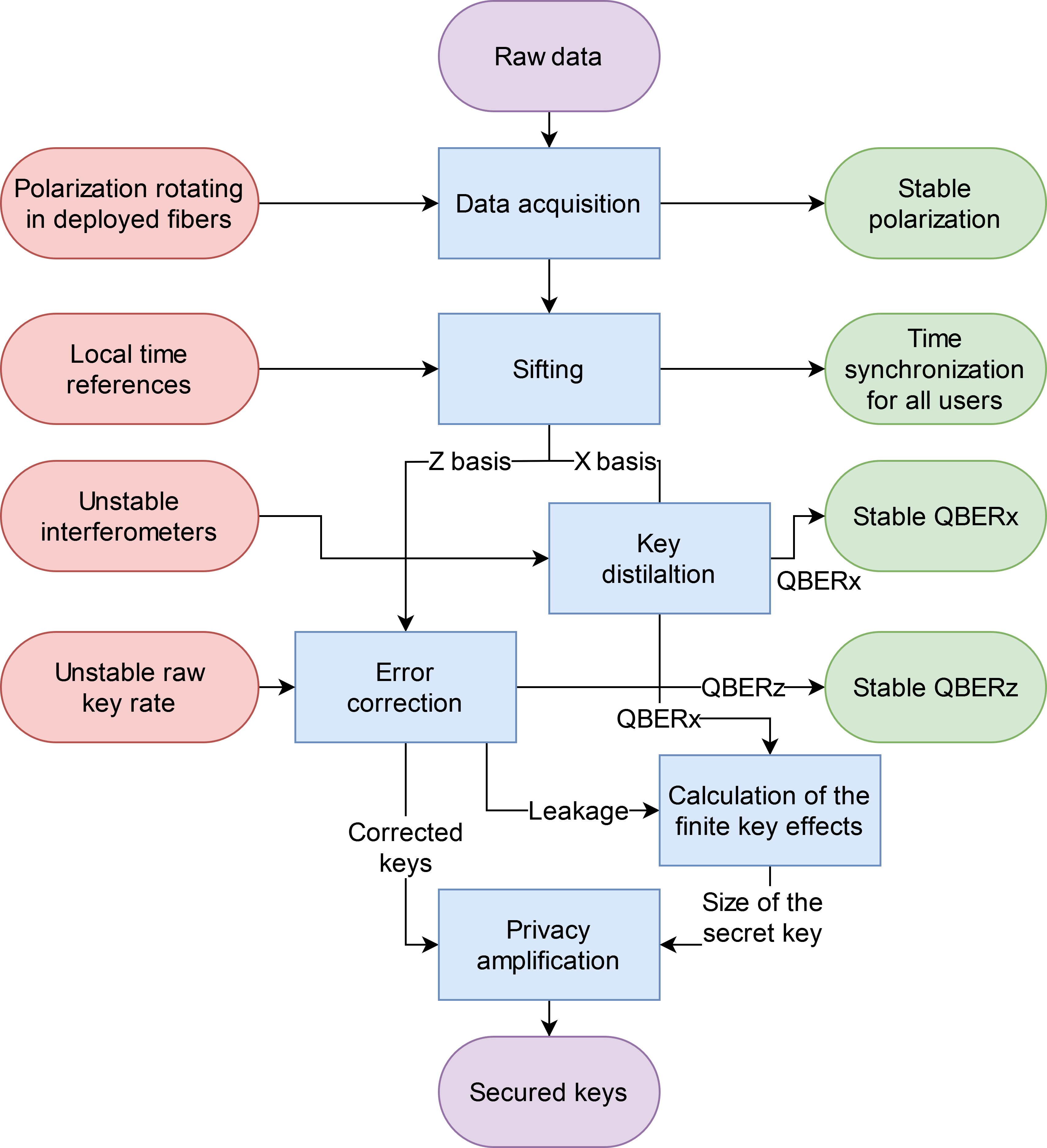}
        \caption{\centering Diagram of the interconnections between the different post-processing steps (blue) required to secure the keys (purple), to address the experimental issues to handle (red), and to perform the stabilization procedures accordingly (green).}
    \label{fig_diagramme_prog}
    \end{center}
\end{figure}

As mentioned above, all the post-processing steps required to extract secure keys are merged into a single software, which receives the raw data transmitted by Alice's and Bob's TDCs and transforms them into secret keys, stored locally on each user's computer.
In addition to key processing, the software enables real-time monitoring of several key QKD indicators, including the QBER in each basis, the number of photons detected, and the time drift between Alice's and Bob's local clocks.
As shown in FIG.~\ref{fig_diagramme_prog}, each of these can be linked to a device that can be controlled to optimize the secure key rate (SKR).
This allows us to continuously correct the instabilities encountered by the various components of the source and analysis modules, which are detailed below.

\subsection{Polarization stabilization}

\begin{figure}[h]
	\begin{center}
   \includegraphics[width=0.95\linewidth]{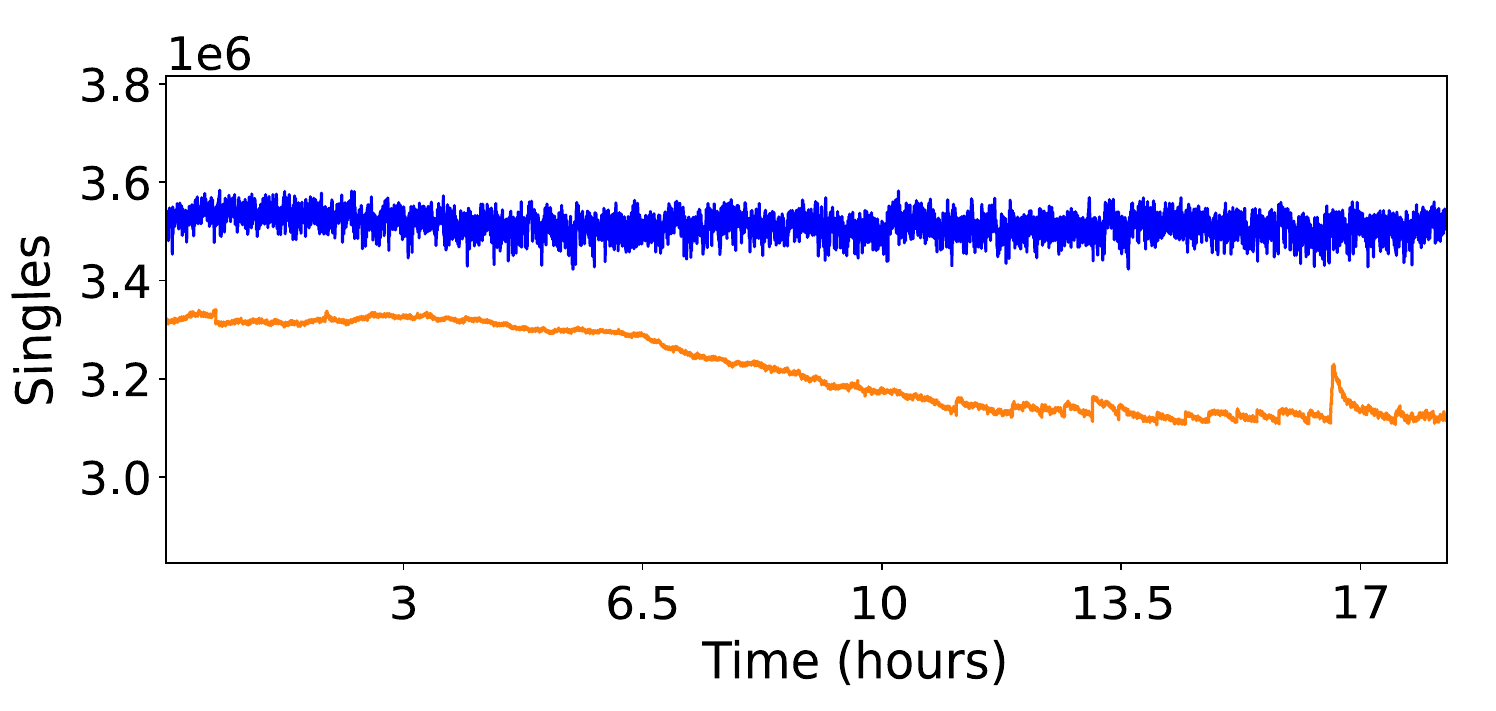}
	\caption{\centering Time trace of the single photon detection rate at Alice's Z basis detector with polarization stabilization (blue curve) and without polarization stabilization (orange curve). The detection rate for the polarization compensated case is maintained stable for more than 17\,h.}
	\label{fig_polar}
    \end{center}
\end{figure}

The first stabilization handled by our software aims at compensating the polarization rotation induced by the deployed fibers.
Despite being mostly underground, the fibers experience significant changes in temperature from day/night cycles, and weather conditions on a daily basis. These induce polarization rotation of the photons measured at the end stations which does not prevent the generation of secret keys, our analyzers being insensitive to polarization variations, however, it can affect the detection efficiency since our SNSPDs are polarization sensitive.
This would impact the global loss budget of the link (up to $6$ additional dB for polarization orthogonal to the optimal). Therefore, to ensure constant maximum detection efficiency, a polarization stabilization protocol must be implemented. We have measured a typical evolution time of several hours as reported on FIG.~\ref{fig_polar}, compatible with the study presented in Ref.~\cite{Qotham_24}.

Since polarization rotations occur at low speed, we add thermal-based fibered polarization controllers, one between the source and each user, which update the polarization every second, to maximize the number of detections on each detector.
It induces a small modulation ($3$\,\%) in the number of detected photons at around $1$\,Hz but fully compensates long-term polarization drifts and improves the mean value of the detection rate, as shown in FIG.~\ref{fig_polar}.

\subsection{Synchronization}

While polarization is an optional optimization for the SKR, some stabilizations are mandatory to perform the QKD protocol.
The first parameter to control is the time synchronization of the different end stations.
In order to monitor coincidences when sharing photon pairs between different users, the local clocks of each TDCs have to stay synchronized.
It is usually done by sharing a dedicated synchronization signal, which can either be transmitted over the fiber but add noise in the detections~\cite{bacco2019field}, or can be sent with a dedicated fiber, which increases the complexity of the QKD infrastructure~\cite{chen_integrated_2021}.

\begin{figure}[htb]
	\begin{center}
   \includegraphics[width=0.95\linewidth]{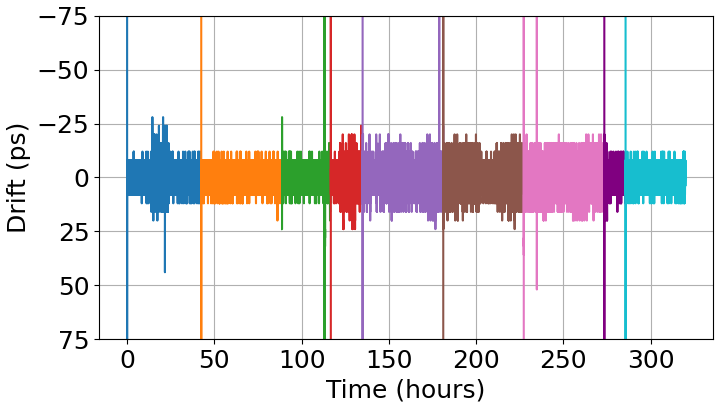}
	\caption{\centering Drift of Bob's rubidium clock compared to Alice's (in ps) as a function of time. The drift mostly stays under $12$\,ps, and can go up to $40$ while under environmental perturbations. Several points go above this limit, due to either a reboot of the QKD program after an evaporation cycle of the SNSPDs, or simply to a bug in the data acquisition, causing a momentary loss of the position of the peak.}
	\label{drift}
    \end{center}
\end{figure}

We implement a protocol, described in Ref.~\cite{synchro_pelet}, to maintain temporal drift between the two users at less than $12$\,ps at all time scales when the analyzers are in a stable environment, and at less than $40$\,ps, even with external perturbations as shown in FIG.~\ref{drift}.
To reach this number, we compute a coincidence histogram between Alice and Bob every 500\,ms, and measure, for each set of data, the position of the coincidence peak (the time delay between the detections of the photons from the same pair at each end station) with a precision of $4$\,ps.
By comparing the position of the peak in two consecutive histograms, we can measure the temporal drift accumulated by the clocks, and then dynamically correct it.
This correction also allows to gradually adapt Alice’s clock frequency to that of Bob in a closer-and-closer manner, thus reducing the passive temporal drift.
This protocol, however, only works with clocks showing a certain degree of passive stability. 
For example, since we use a $120$\,ps wide coincidence window to discriminate photon pairs (the coincidence peak) from photonic noise, the accumulated clock's drift between two consecutive histograms has to be smaller than 120\,ps. 
Built-in quartz clocks in the TDCs show an average drift of a few ns/s, which is too large for our purpose, as we compute histograms only twice per second.
They have been therefore replaced by rubidium atomic clocks, showing drifts of $7$\,ps/s, which allow to obtain the results described in FIG.~\ref{drift}.

Let us emphasize that our synchronization protocol uses the same entangled photons as the one used for generating the secret key. As a matter of fact, the data used for synchronization does not reveal any information about the raw key. As a consequence, our protocol can simultaneously generate raw keys and evaluate the synchronization without lowering the SKR or increasing the QBER. The advantage of such a configuration is that it eliminates the need for an additional fibre to transmit a clock signal while it does not impact the key rate.
This makes our solution ideal for entanglement-based communication at a metropolitan scale (links shorter than $100$\,km, or $35$\,dB).
This protocol, however, shows limitations for higher losses. 
More precisely, a minimal amount of data is required to build a coincidence histogram with sufficient SNR, which increases the integration time required when the losses get higher.
Therefore, the maximum losses that can be handled by our protocol are directly linked to the passive stability of the clocks, see discussion section~\ref{distance}. For instance, we have been limited to 55\,dB for rubidium clocks, but cesium clocks, that are ten times more stable, could theoretically increase this number up to $65$\,dB.

\subsection{QBERz stabilization} 

One of the main advantages of using time as an observable is that errors in the Z basis are mostly due to the detections of double pairs.
This type of event occurs when two different photon pairs are generated in less than $120$\,ps (the size of our coincidence window), which could result in Bob measuring a photon from the first pair, and Alice a photon from the second, while still considering this joint measurement as a coincidence.
However, since the detected photons are not correlated, these double pairs have a $50$\,\% chance of generating an error and must therefore be minimized.
Given that photon pair generation follows Poissonian statistics, the number of double pair generated increases linearly with the pump power.
Consequently, while an increase in pump power enhances the rate of photon pair generation and therefore improves the SKR, it also raises the proportion of double-pair, thereby elevating the QBER and potentially reducing the overall SKR.

\begin{figure}[h]
	\begin{center}
\includegraphics[width=0.95\linewidth]{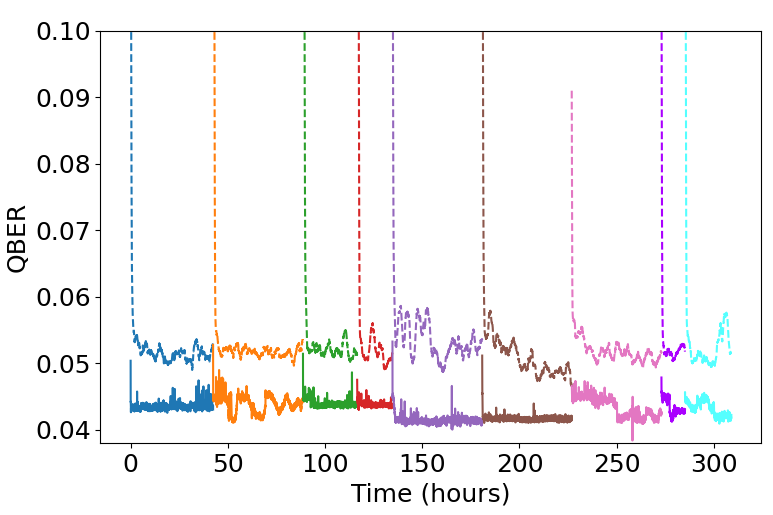}
	\caption{\centering Stability over time of the QBER in the Z and X basis, measured during a single QKD run. The solid and dashed lines represent QBERz and QBERx respectively. In our experiment, the optimal mean number of photon pairs per coincidence windows leads the QBERz to reach an average value of $4.3$\,\%, while the QBERx reaches an average of $5.4$\,\%.}
	\label{fig_QBERz}
    \end{center}
\end{figure}

We study both theoretically and experimentally the trade-off between these two effects for our loss budget, by simulating not only the quantum state of the pairs after transmission through the link and analyzers, but also by taking into account the post-treatment as described in Ref.~\cite{cai2009finite} and reported in Ref.~\cite{pelet2023operational}.
We find an optimal value of $4.3$\,\% for the QBER in the Z basis (QBERz), which represents the maximum number of double pairs we can tolerate before the number of errors becomes too high and compromises the SNR.

This study shows that, to continuously ensure a maximum key rate, we need to stabilize the QBERz, which requires a stable photon-pair flux.
A variable optical attenuator is therefore placed between the pump laser and the non-linear crystal in the source (see FIG.~\ref{setup}), with a feedback control from the QBERz.
If the error rate is too high, the attenuation increases, and if it is too low, it decreases, leading to the stability depicted in FIG.~\ref{fig_QBERz}.

\subsection{QBERx stabilization} 

Measurements in the X basis rely on non-local interference~\cite{Brendel_1992}, which requires the use of two identical interferometers, one for each user.
This condition implies to build two interferometers with identical delays, but more importantly, to lock their relative phase at all times.

\begin{figure}
    \centering
    \includegraphics[width=0.8\linewidth]{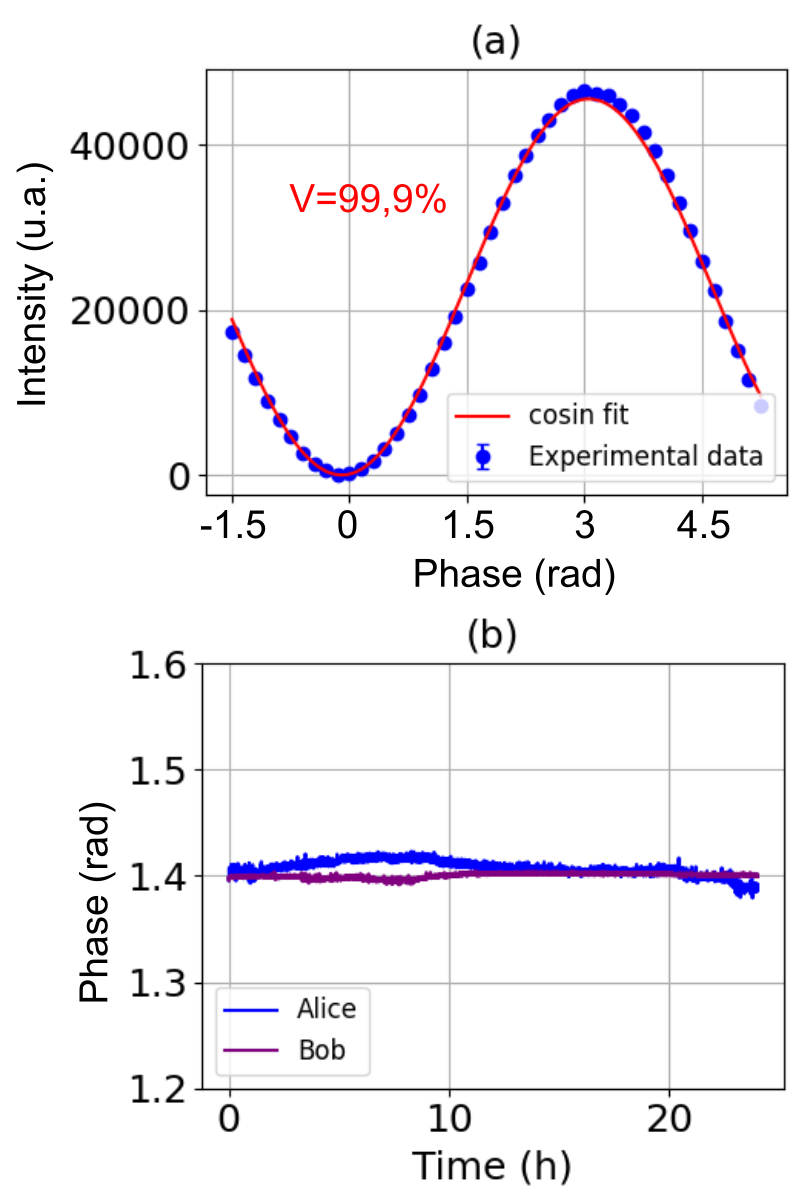}
    \caption{\centering
    Classical characterization of Exail's delay line interferometers.
    (a) Raw visibility (blue) with a cosine fit (red).
    (b) Phase stability of Alice’s (blue) and Bob’s (purple) interferometers. Both measurements have been performed with an attenuated continuous-wave laser at 1560\,nm; phases are shown as a function of time. Data are acquired with a 1\,s sampling interval at the most phase-sensitive operating point.}
    \label{carac2}
\end{figure}

As discussed above, our setup only uses quantum resources for stabilization and synchronization~\cite{pelet2023operational}. Beyond a direct influence on the SKR, losses have, in addition, a tremendous effect on our system and one of the major challenges for reaching long distance QKD is to improve the performance of our analyzers previously used in~\cite{pelet2023operational} in terms of losses and stability. We have used ultra-stable delay line interferometers from Exail. Not only they exhibit high-visibility single-photon interference, as presented in FIG.~\ref{carac2}a, but they also show low-loss operation as reported in table~\ref{carac}. In addition, they report an exceptional passive phase stability since, over 24 hours, we measured a mean phase variation of $10^{-7}$\,rad/s as presented in FIG.~\ref{carac2}b in laboratory condition, i.e. daily temperature fluctuation of $\pm1^\circ$\,C.
Concerning the Franson-type interference, they have shown identical FSR of $2.5$\,GHz to better than $10^4$\,Hz, validating the non-local Franson condition ($\delta L\ll \mathcal{L}_{coherence}^{single\;photon}$). We also measured, under laboratory conditions, a non-local two-photon interference associated with a visibility of 99.7\,\% as reported in Table~\ref{carac}.
\begin{table}
    \centering
    \begin{tabular}{|c|c|c|c|c|}
        \hline
        UMZI & \textbf{Stability} & \textbf{FSR} & \textbf{Two-photon} & \textbf{Losses} \\
        & & & \textbf{visibility} &\\
             & (rad/s)            & (GHz)        & (Raw / Net)           & (dB) \\
        \hline
        Alice & 6.5(6)$\times$\,$10^{-7}$ & 2.51310(2) & \multirow{2}{*}{0.997(8) / 1.000(2)} & 0.6(1) \\
        \cline{1-3} \cline{5-5}
        Bob   & 4.7(5)$\times$\,$10^{-7}$ & 2.51309(1) & & 0.5(1) \\
        \hline
    \end{tabular}
    \caption{\centering Characterization of the interferometers used for Alice and Bob analyzers.}
    \label{carac}
\end{table}

Still a piezoelectric phase control is installed on Alice's interferometer, allowing to actively control its phase to follow Bob's natural drift.
It stretches to always minimize the QBER in X basis (QBERx) for the whole duration of the key sharing process, reaching an average of $5.4$\,\% in our experiment.
This process directly uses the QBERx as the error-target to minimize, which allows this stabilization to rely only on the quantum data, therefore avoiding the use of a dedicated classical signal, that should be sent in each device.
This is therefore an ideal process for QKD since it does not require any resource overhead and does not add any optical noise to our measurement.
However, a minimum amount of data is required to get enough statistics to compute each value of QBERx measured and, depending on the link losses, the integration time needed spans from few milliseconds to several minutes.

\section{Results}

\subsection{2 users at 50\,km over 325\,hours}

\begin{figure}[h]
	\begin{center}
   \includegraphics[width=0.95\linewidth]{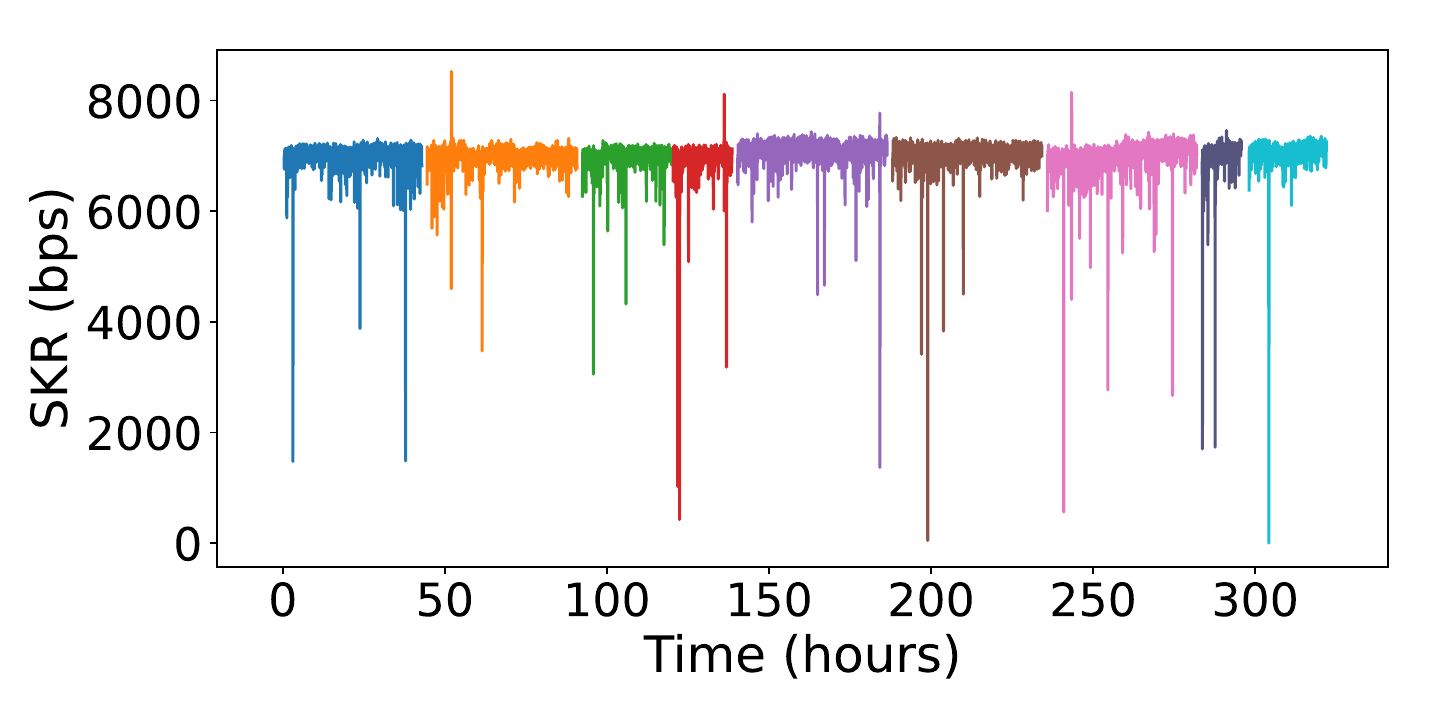}
	\caption{\centering Evolution of the secret key rate (SKR) over time. Each color represents an automated restart of the program after an evaporation cycle of the SNSPDs, or a crash of the computers, or of the classical communications.}
	\label{fig_300}
    \end{center}
\end{figure}

To illustrate the stability of our system, we perform a 325\,hours-long QKD experiment which is, to the best of our knowledge, the longest run performed to date in real-field conditions using entanglement.
During this time, no human intervention has been required, and everything from the generation of the key to the error management has been handled automatically.
The length of the experiment required the SNSPDs to perform $7$ evaporation cycles (one cycle of $1$\,hour every $48$\,hours usually), after each of which the key sharing process restarted automatically.
Each of those cycles is started automatically when their internal temperature starts to rise naturally, and the QKD resumes as soon as the temperature goes back under the operating threshold of our detectors ($\sim$0.8\,K).
As shown in FIG.~\ref{fig_300}, we measure an average key rate of $\sim$7.069\,kbps, which takes into account the evaporation cycles, that lower the key rate by $\sim$\,1\% compared to an uninterrupted experiment by inducing downtimes in the key generation (green point of FIG.~\ref{skr_loss}).

Because of the evaporation cycles, one software failure at $t=125$\,h, and one cryostat failure at $t = 280$\,h that forced an early evaporation, the program had to restart the whole QKD protocol in total $9$ times during the experiment. 
Each restart requires to restabilize the phase of the interferometer, which usually takes around one minute, during which the average QBERx is higher and the key rate lower.
In an $325$-hours-long test, $9$ minutes of restabilization for the phase is negligible, but for higher distances, longer times would be needed between each computation of the QBERx since we observed that the first stabilization can take up to 20\,minutes.

\subsection{2 users at 100\,km over 30\,h}\label{distance}

\begin{figure}[h]
    \centering
    \includegraphics[width=1\linewidth]{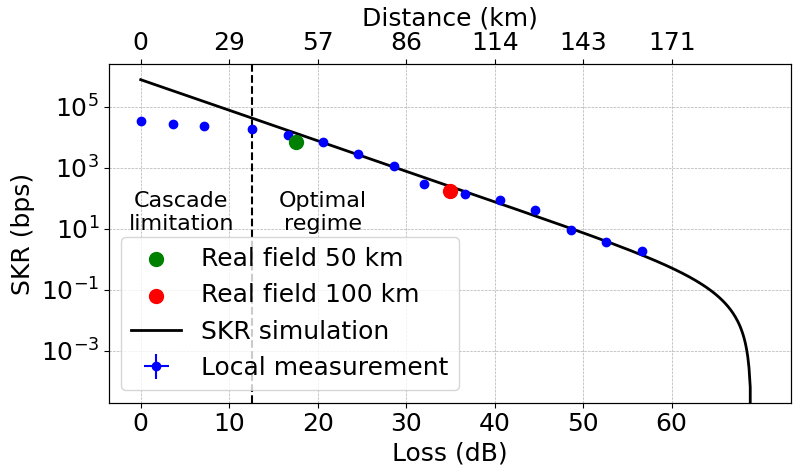}
    \caption{\centering SKR as a function of transmission losses in the quantum channel and distance, assuming a fiber attenuation of 0.35\,dB/km. The dotted line separates two regions: one where the Cascade protocol does not yield the optimal SKR, and one where the maximum SKR can be achieved. The black line shows the simulated SKR curve using our experimental parameters; blue points correspond to SKR measurements for given attenuations, while the green and red points indicate measurements obtained in real field scenario at 50\,km and 100\,km, respectively.}
    \label{skr_loss}
\end{figure}

\begin{figure}
    \centering
    \includegraphics[width=0.98\linewidth]{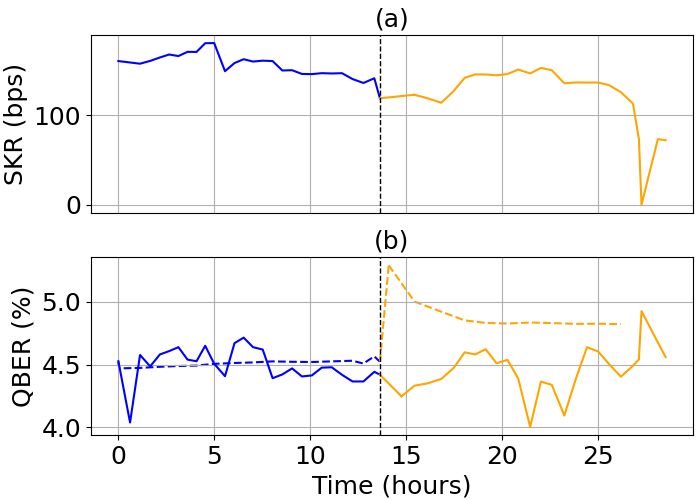}
    \caption{\centering The 30-hour measurement was deployed outside the laboratory over a distance of 100\,km. The measurements were carried out in $2\times15$-hour sessions (blue and orange), represented by the dotted line, which delineates the measurement cycle of the SNSPDs. (a) : SKR as a function of time. (b) : full and dash curve are respectively the QBERz and QBERx as a function of the time.}
    \label{100km}
\end{figure}

As discussed earlier, the optimal SKR corresponds to a sweet spot that depends on the transmission losses and a relative mean number of photon pairs per second. We report in FIG.~\ref{skr_loss} the theoretical estimation and the experimental value of the SKR for various distances. When we are in a low-loss regime (below 13\,dB), we observe that the optimal SKR is limited by our computing power for error correction since our cascade~\cite{martinez2014demystifying} implementation limits the processed SKR due to $\sim$600 sequential instances that cannot be performed simultaneously.

In high-loss regime, we demonstrate a secret key rate of 1.8\,bps up to a total loss of 56.6\,dB in the quantum channel. It is interesting to note that the limitation arises from the synchronization procedure that fails to estimate in reasonable time the feedback to the rubidium clocks.

On the real-field scenario, for practical loss of 33.5\,dB corresponding to 100\,km, our system has delivered an SKR of up to 175\,bps (red dot on FIG.~\ref{skr_loss}). 

\subsection{36 users at 50\,km over 1\,h}

\begin{figure}[h]
	\begin{center}
   \includegraphics[width=1\linewidth]{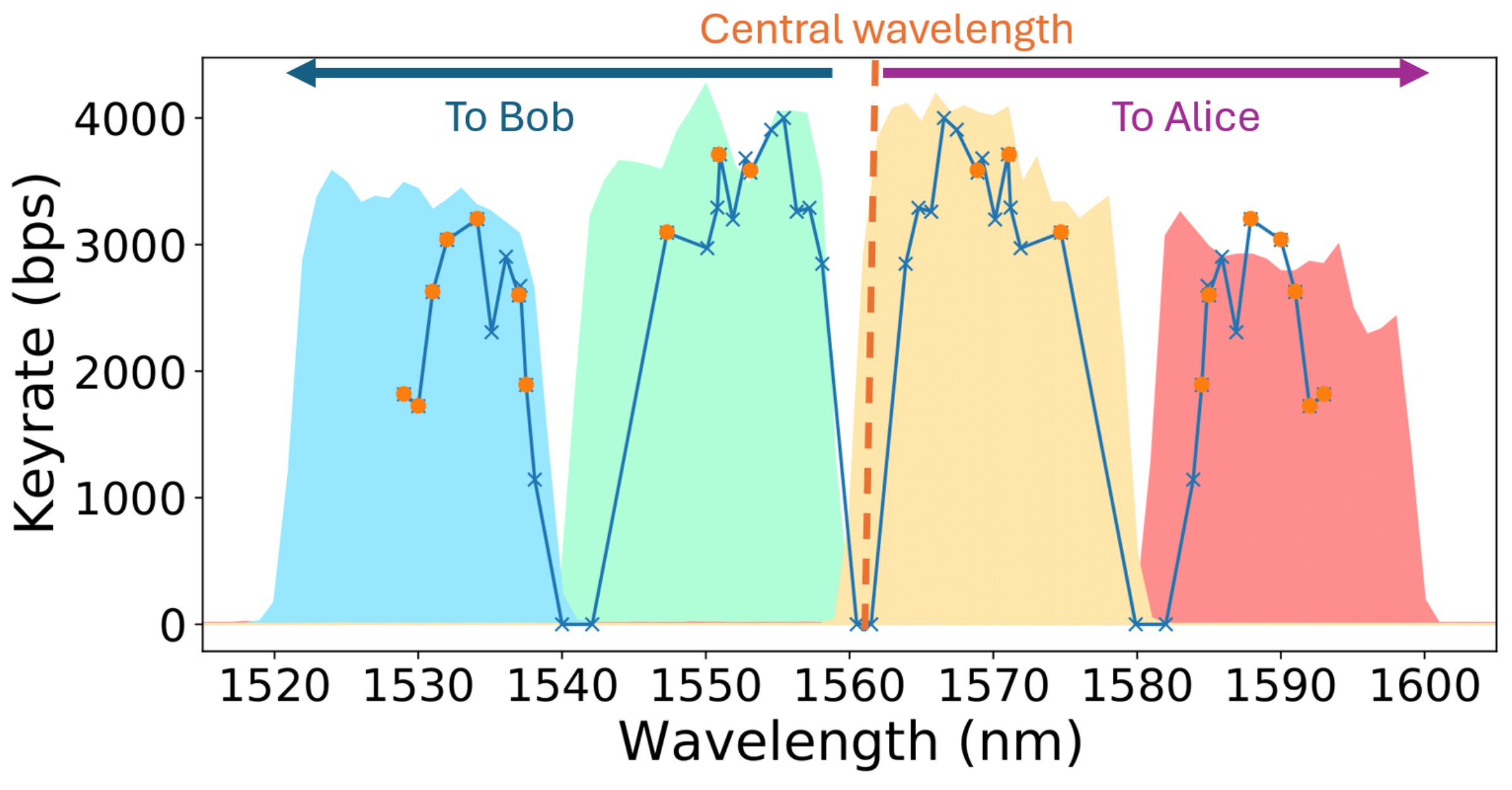}
	\caption{ \centering Evolution of the secure key rate as a function of the wavelength of the DWDM used. Orange dots are measurements performed in real-field while the blue crosses are performed in the lab for equivalent attenuation ($\sim$20\,dB total attenuation). The colored background represents the measured CWDM transmission used to split the spectrum of the source between Alice and Bob.}
	\label{fig_wavelength}
    \end{center}
\end{figure}

Our source is designed to generate photon pairs over a wide spectrum ($80$\,nm at FWHM), allowing theoretically to extract up to $40$ independent pairs of 100\,GHz-quantum channels.

Building a QKD network exploiting all those channels simultaneously would require an industrial implementation with 40 analyzers and up to 160 SNSPDs, which is expensive and challenging in a lab environment.
More realistically, we can get close to actual network conditions by making all 40 pairs of channels simultaneously available using two DWDM with 40 outputs, and then perform QKD on each pairs sequentially.
All the results are gathered in FIG.~\ref{fig_wavelength} and show that a wide portion of the spectrum is usable, and therefore that our broadband source is suitable for real field quantum network implementations.

However, we can see on FIG.~\ref{fig_wavelength} that no secret key is generated near $1561$\,nm and $1581/1541$\,nm.
This is due to the CWDMs edges that do not overlap completely.
They create holes in the transmission profile of the setup, leading to a decrease and extinction of the SKR in some regions.
This could be avoided by using an active wavelength selective switch (WSS) as in Ref.~\cite{WSS,WSS2} instead of the passive 18\,nm CWDM used, allowing full access to the whole available spectrum.

\section{Conclusion}
In this paper, we have demonstrated the long-term stability of an entanglement-based QKD setup, deployed in real-field over $50$\,km of optical fibers.
Every stabilization, synchronization, and post-processing operations are handled continuously and automatically thanks to an in-house software, allowing our system to be fully autonomous.
We demonstrate this automation over $325$\,hours of key, with an average secure key rate of $7.069$\,kbps.

We also demonstrate the adaptability of our setup for different distances of communication, up to 100\,km (i.e. $56$\,dB of total transmission losses), as well as the resilience of our synchronization protocol to strong environmental perturbations.
Lastly, we have studied the feasibility of QKD exploiting the complete spectrum of the photon pairs generated by our source.
We show that up to 18 pairs of quantum channels can be exploited, creating as much as independent QKD links.
With this, we could investigate various quantum network topologies over a metropolitan area, with simultaneous key sharing between all connected users, and no trusted nodes.
Such a stable QKD network already serves as a testbed for more advanced quantum communication protocols, which represents a new step towards large-scale quantum networks.

\section{Acknowledgements}
This work has been conducted within the framework of the French government financial support managed by the Agence Nationale de la Recherche (ANR), within its Investments for the Future programme, under the Université Côte d'Azur UCA-JEDI project (Quantum@UCA, No. ANR-15-IDEX-01), and under the Stratégie Nationale Quantique through the PEPR QCOMMTESTBED project (No. ANR 22-PETQ-0011). This work has also been conducted within the framework of the OPTIMAL project, funded by the European Union and the Conseil Régional SUD-PACA by means of the “Fonds Européens de développement regional” (FEDER). The authors also acknowledge financial support from the European Commission, through the Project Nos. 101114043-QSNP and 101091675-FranceQCI. The authors are grateful to S. Canard, A. Ouorou, L. Chotard, L. Londeix, and to Orange for the installation and the connection of the dark fibers between the three different sites of our network, as well as for all the support they provided for their characterization. The authors also thank the Métropole Nice Côte d'Azur, the Observatoire de la Côte d’Azur (OCA) and the Inria Centre at Université Côte d'Azur for the access to their buildings and for their continuous help in making this network a reality. The authors also acknowledge IDQuantique, Exail and Swabian Instruments GmbH teams for all the technical support and the development of new features that were needed for the implementation of our operational QKD system and related experiments.

N.L. would like to express his gratitude to EUR SPECTRUM and PEPR Qcommtestbed 22-PETQ-0011 for their financial support for his thesis.

Y.P. would like to express his gratitude to Accenture for their financial support for his thesis.

\bibliography{apssamp}

\end{document}